\documentclass[twocolumn,aps,prl,superscriptaddress,preprintnumbers,nofootinbib]{revtex4-2}
\usepackage{graphicx}
\usepackage{subcaption}
\usepackage{color}
\graphicspath{{figures/}{fig/}}
\usepackage{amsmath}
\usepackage{amssymb}
\usepackage{bm}
\usepackage{slashed}
\usepackage{epsfig}
\usepackage{amsfonts}
\usepackage{epstopdf}
\usepackage{hyperref}
\usepackage{bbm}
\usepackage{textcomp}
\usepackage{color}
\usepackage{dsfont}

\newcommand{\sect}[1]{{\it \textbf{#1.} --- }}

\def\beq{\begin{equation}}
\def\eeq{\end{equation}}
\def\bea{\begin{eqnarray}}
\def\eea{\end{eqnarray}}
\begin{document}
\preprint{SLAC-PUB-250331, ZU-TH 24/25}

\title{Three-Loop QCD corrections to the production of a Higgs boson and a Jet}

\author{Xiang Chen}
\email{xiang.chen@physik.uzh.ch}
\affiliation{Physik-Institut, Universit\"{a}t Z\"{u}rich, Winterthurerstrasse 190, CH-8057 Z\"{u}rich, Switzerland}

\author{Xin Guan}
\email{guanxin@slac.stanford.edu}
\affiliation{SLAC National Accelerator Laboratory, Stanford University, Stanford, CA 94039, USA}

\author{Bernhard Mistlberger,}
\email{bernhard.mistlberger@gmail.com}
\affiliation{SLAC National Accelerator Laboratory, Stanford University, Stanford, CA 94039, USA}

\date{\today}

\begin{abstract}
We compute three-loop QCD corrections to the scattering amplitude of a Higgs boson and three partons interfered with its tree level counterpart.
Specifically, we derive our results in the generalized leading color limit. 
Our results are represented in terms of so-called multiple polylogarithms and thus ready for use in phenomenological predictions. 
We provide the three-loop amplitudes necessary to compute the production cross section of a Higgs boson and a hadronic jet at hadron colliders and the decay probability of a Higgs boson to three hadronic jets at lepton colliders.
\end{abstract}

\maketitle
\allowdisplaybreaks

\section{Introduction}
\label{sec:introduction}
The discovery of the Higgs boson by the ATLAS~\cite{ATLAS:2012yve} and CMS~\cite{CMS:2012qbp} experiment at the Large Hadron Collider (LHC) ushered in a new era of exploration of fundamental forces. 
For the first time we gained experimental access to probe the interactions of all fundamental particles of the Standard Model and thus a brand new opportunity to test the foundations of our understanding of nature.
Today, we explore the rich implications of the Higgs boson by performing high precision comparison of experimental data with theoretical predictions for the outcome of LHC scattering experiments. 
The large data sets already collected by the LHC and in particular the advent of the upcoming High-Luminosity Phase~\cite{ZurbanoFernandez:2020cco} enable a precision phenomenology program testing the Standard Model and searching for new physics beyond this paradigm.
The high experimental precision must be matched by equally precise theoretical predictions based on the cutting edge of our understanding of Quantum Field Theory (QFT). 
In this Letter, we achieve a new milestone of precision calculations at the core of such predictions.

Precision computations for the LHC are performed exploiting the factorization of high energetic scattering processes of the QCD partons - the constituents of the two scattering protons - and the low-energy dynamics of the individual protons. 
Within this framework, we compute scattering cross sections based on the first principle Lagrangian of the Standard Model or other theories of nature using perturbative QFT.
Precision is achieved by computing cross sections to higher and higher power in an expansion in coupling constants. 
The largest of all coupling constants at LHC energies is the coupling of the strong force $\alpha_S$. 
Achieving high order results in the expansion in powers of $\alpha_S$ is a formidable task challenging our technical and theoretical abilities of using QFT.

An overwhelming number of about ninety percent of all Higgs bosons at the LHC are produced via the gluon fusion mechanism.
This mechanism describes the fusion of two initial state gluons via a virtual top-quark loop which in turn radiates the Higgs boson. 
Perturbative QCD corrections for this production mechnism are notoriously large and consequently very high order predictions are required to rival experimental precision. 
A powerful approach is the construction of an effective field theory~\cite{Spiridonov:1988md,Inami1983,Shifman1978,Wilczek1977} integrating out the degrees of freedom of the top quark yielding the first term in an expansion in inverse powers of the large top quark mass. 
This approach reduces the top-quark loop at Born level to a point-like interaction coupling the Higgs field $H$ directly to the gluon field strength $G_a^{\mu\nu}$ via a Wilson coefficient~\cite{Chetyrkin:2005ia,Schroder:2005hy,Chetyrkin:1997un,Kramer:1996iq,Kataev:1981gr} $C^0$.
\begin{equation}\label{eq:heft}
    \mathcal{L}_{eff}=\mathcal{L}_{SM,5}-\dfrac{C^0}{4}HG_{a}^{\mu\nu}G_{a,\mu\nu}.
\end{equation}
 \begin{figure}[!t]
  \begin{center}
\vspace{0.3 cm}
\includegraphics[scale=.45]{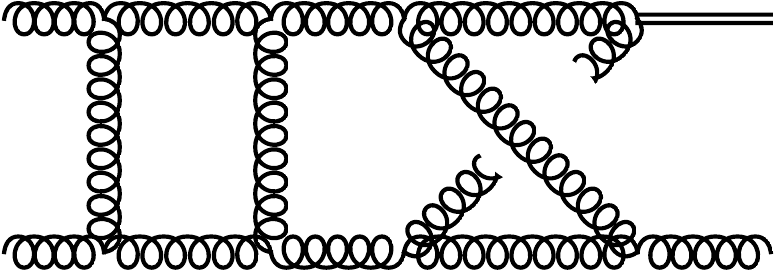}
  \end{center}
  \caption{\label{fig:examplediag}Example diagram for the leading-color, three-loop scattering amplitude of three gluons and a Higgs boson.
   }
  \vspace{-0.5cm}
  \end{figure}
The application of this effective field theory has enabled frontier computations of the inclusive Higgs production cross section through next-to-next-to-next-to-leading order (N$^3$LO) in QCD perturbation theory~\cite{Mistlberger:2018etf,Anastasiou:2016cez,Ravindran:2003um,Anastasiou2002,Harlander:2002wh,Dawson:1990zj} and predictions including the massive top quark were obtained already through next-to-next-to-leading order (NNLO)~\cite{Niggetiedt:2024nmp,Czakon:2024ywb,Spira:1995rr,Graudenz:1992pv,Caola:2015wna}.
Higgs phenomenology is however much richer than only inclusive rates and a vast range of observables differential in the decay products of the Higgs boson allow us to study this new particle in great detail. 
The advances in our understanding of differential cross section of the past two decades enabled the computation of fully differential predictions for the production of a Higgs boson through N$^3$LO in QCD perturbation theory~\cite{Boughezal2016,Anastasiou2005,Catani:2007vq,Chen:2022xnd,Cieri:2018oms,Dulat:2018bfe,Chen:2019fhs}.
A leap in our ability to perform differential computations with hadronic jets occurred about one decade ago~\cite{Czakon:2011ve,Boughezal:2011jf,Cacciari:2015jma,GehrmannDeRidder:2005cm,Daleo:2006xa,Currie:2013vh,Boughezal:2015eha,Gaunt:2015pea} leading to fully differential predictions of the gluon fusion production of a Higgs boson in association with a jet at NNLO~\cite{Boughezal2015b,Boughezal:2015aha,Chen2015,Chen2016}. 
The additional final state jet is a requirement for any observable dependent on the Higgs boson having non-vanishing transverse momentum and enables to further broaden the set of observables of Higgs boson physics.
Extending predictions for the production of a Higgs boson in association with a jet to N$^3$LO remains a formidable challenge.
Motivated by the ambition to enable one-percent precision phenomenology at the LHC~\cite{Caola:2022ayt} we take here a decisive step towards this goal.

An essential ingredient of perturbative corrections is the availability of scattering amplitudes to the required order in perturbation theory. In this Letter, we focus on three-loop corrections to amplitudes describing the scattering of a Higgs boson and three partons. Results at one and two loops for these scattering amplitudes were obtained in refs.~\cite{Badger:2006us,Badger:2009hw,Badger:2009vh,Dixon:2009uk,Gehrmann:2011aa,Gehrmann:2023etk}. 
The past computations of these scattering amplitudes have provided not only key ingredients to today's LHC phenomenology program but also led to deeper understanding of the functions these amplitudes are comprised of (see for example refs.~\cite{Gehrmann:2001jv,Gehrmann:2024tds}), to improved computational techniques~\cite{Peraro:2020sfm} and to a flurry of activity of understanding formal gauge theory~\cite{Brandhuber:2012vm,Dixon:2020bbt,Dixon:2022rse,Lin:2021qol,Dixon:2022rse,Dixon:2023kop}. 
The technology enabling the computation of these particular scattering amplitudes is then applicable to a range of scattering processes, most notably the scattering amplitudes relating three partons and a vector boson~\cite{Gehrmann:2013vga,Gehrmann:2022vuk,Gehrmann:2023jyv,Gehrmann:2023zpz}. 

Figure~\ref{fig:examplediag} shows an example Feynman diagram contributing to the three-loop corrections of the scattering amplitude relating a Higgs boson and three gluons.
To tame the immense complexity of these amplitudes we restrict our computation to the generalized leading color contributions treating the number of colors $N_c$ of QCD and number of light quark flavors $n_f$ as large and neglect corrections that are suppressed beyond the leading term in these parameters by inverse powers of the number of colors.
Due to the direct coupling of the Higgs boson to gluons in the effective theory of eq.~\eqref{eq:heft} and the color-singlet nature of the Higgs the generalized leading color limit of the scattering amplitude is not only given by so-called planar diagrams~\cite{tHooft:1973alw}. Non-planar diagrams, such as the one displayed in fig.~\ref{fig:examplediag}, drastically increase the complexity of perturbative computations and put our results at the forefront of the field.

Our results will find application beyond the prediction of the production cross section for a Higgs boson and a jet. First, our scattering amplitudes are a natural ingredient to the inclusive production probability of a Higgs boson at the next order, N$^4$LO. Furthermore, the advent of future lepton collider facilities promise to study hadronic decays of the Higgs in astounding detail. A range of planned facilities like the FCC-ee~\cite{FCC:2018byv,FCC:2018evy}, CEPC~\cite{CEPCStudyGroup:2018ghi}, ILC~\cite{Behnke:2013lya}, $C^3$~\cite{Vernieri:2022fae} or future muon colliders would provide the opportunity to study jets as decay products of the Higgs boson. 
While inclusive hadronic decay rates are known to be of astounding fourth order in QCD~\cite{Baikov:2005rw,Baikov:2006ch,Davies:2017xsp,Herzog:2017dtz}, differential two and three jets rates are currently available through N$^3$LO~\cite{Mondini:2019gid} and NNLO~\cite{Mondini:2019vub,Fox:2025cuz} respectively. Advancing the latter to N$^3$LO requires the computation of the scattering amplitudes computed here.

\section{Set-Up}
\label{sec:set-up}

We compute three-loop scattering amplitudes relating a Higgs boson $H$ with momentum $q$ and three partons with momenta $k_i$, $i\in \{1,2,3\}$.
We start by considering the {\it decay region}, which describes the process of the Higgs boson decaying to three gluons or to a gluon and a quark anti-quark pair.
\bea
    H(q)&\to& g(k_1)+g(k_2)+g(k_3), \nonumber\\
    H(q) &\to& g(k_1)+q(k_2)+ \bar q(k_3).
\eea
The scattering amplitude $\mathcal{A}_X$ for the process $X$ (e.g. $X=H\to ggg$) can then be expanded in powers of the bare strong coupling constant $\alpha_S=\frac{g_S^2}{4\pi}$,
as
\beq
\mathcal{A}_X(\alpha_S,C^0)=g_S C^0\sum_{n=0}^\infty a_S^n \mathcal{A}_X^{(n)},\hspace{1cm}
a_S=\frac{\alpha_S}{\pi} N_\epsilon^{-1},
\eeq
where $N_\epsilon=e^{\epsilon \gamma_E}(4\pi)^{-\epsilon}$.Throughout this Letter we work in the framework of dimensional regularization with $d=4-2\epsilon$ space time dimensions in the modified minimal subtraction scheme $\overline{\text{MS}}$ to regulate infrared and ultraviolet singularities of our scattering amplitudes. The definition of $a_S$ in the above equation anticipates renormalization in the $\overline{\text{MS}}$ by absorbing universal factors of $(4\pi)$ and the Euler-Mascheroni constant $\gamma_E$. To compute the ingredient necessary for physical scattering cross sections we multiply our scattering amplitudes with their tree level counterparts and sum over spin and color degrees of freedom of the external partons.
We denote the resulting interfered matrix elements by
\beq
\mathcal{M}_X^{(n)}=\sum\limits_{\text{spin, color}}\mathcal{A}_X^{(n)}\mathcal{A}_X^{\dagger,\,(0)}.
\eeq
These matrix elements are scalar functions of the Lorentz invariant scalar products of the external momenta and we choose in particular the parametrization
\beq
s=\frac{(k_1+k_2)^2}{q^2},\hspace{0.5cm}
t=\frac{(k_2+k_3)^2}{q^2},\hspace{0.5cm}
u=\frac{(k_1+k_3)^2}{q^2}.
\eeq
Note, that $k_i^2=0$ and $q^2=m_h^2$ for an on-shell Higgs boson.
The three variables are constrained by the identity
\beq
1=s+t+u,
\eeq
and we choose $t$ and $u$ as our independent variables alongside $q^2$, which carries the energy dimension of the amplitude. In addition, we introduce the variable $Q^2=-q^2- i0$, which a has an infinitesimal imaginary part such that the matrix elements in decay kinematics expressed in terms of $Q^2$, $t$ and $u$ don't have any explicit imaginary parts.
In decay kinematics we find
\beq
s>0,\hspace{1cm}
t>0,\hspace{1cm}
u>0.
\eeq

There are three distinct matrix elements required for the production of a Higgs boson and a parton in the collision of two partons.
We refer to this configuration as {\it scattering region }.
\bea
    g(-k_1)+g(-k_2)\to H(q)+g(k_3), \nonumber\\
    g(-k_1)+q(-k_2)\to H(q)+q(k_3), \nonumber\\
    q(-k_1)+\bar q (-k_2)\to H(q)+g(k_3).
\eea
We obtain the associated matrix elments by crossing the decay matrix elements and permuting of external momenta.
Exchanging quarks and anti-quarks leaves the interfered matrix elements invariant and all other configurations can simply be obtained by re-labeling of the momenta.
In the scattering region, we find that 
\beq
s>0,\hspace{1cm}
t<0,\hspace{1cm}
u<0.
\eeq
For convenience, we introduce the variables $\hat t=-t$ and $\hat u=-u$ with which we express our scattering matrix elements in the scattering region.

The main results of this Letter are the scattering matrix elements $\mathcal{M}_X^{(n)}$ through three loops, $n\in\{0,1,2,3\}$. 
Through two loops we perform our computation for QCD with $n_f$ massless quark flavors and general color gauge group of $SU(N_c)$. In the case of the three-loop amplitudes, we limit our computation to the generalized leading color limit. In this limit we treat the number of colors $N_c$ and flavors $n_f$ as large and neglect contributions suppressed by powers of  $1/ N_c^2$. 
This generalized leading color limit has shown to be a highly effective approximation for similar scattering amplitudes~\cite{Agarwal:2023suw,Gehrmann:2023jyv}, as can be easily understood from the fact that it parametrically approximates matrix elements up to $\mathcal{O}(10\%)$ corrections.
Specifically, this implies that we include the coefficients of the terms proportional to any one of the factors $\{N_c^6, N_c^5n_f, N_c^4n_f^2, N_c^3n_f^3\}$ when all three external partons are gluons and $\{N_c^5, N_c^4n_f, N_c^3n_f^2, N_c^2n_f^3\}$ otherwise. 
Our results are expressed in terms of a Laurent expansion in the dimensional regulator $\epsilon$ including all terms required for a N$^3$LO computation of the discussed scattering and decay processes.
The coefficients of these expansions are expressed in terms of ratios of polynomials and generalized polylogarithms~\cite{Goncharov:2001iea} of our kinematic variables.

\section{Calculation}
\label{sec:calculation}

We generate the Feynman diagrams for our scattering amplitudes with {\tt qgraf}~\cite{qgraf}. We compute gauge invariant scattering amplitudes in conventional dimensional regularization using tensor decomposition (see for example refs.~\cite{Gehrmann:2011aa,Gehrmann:2013vga,Peraro:2020sfm,Goode:2024cfy,Goode:2024mci}) through two loops and interferences of tree and three-loop scattering amplitudes. We use our in-house {\tt Mathematica} and {\tt C++} packages to compute our amplitudes, including contracting the Lorentz and color indices. 
Then, we express all the squared amplitudes as linear combinations of scalar integrals and classify them in terms of integral families.
We utilize the package {\tt CalcLoop}~\cite{calcloop} to fully exploit external symmetry among Feynman integrals.
Once we identified a minimal set of Feynman integrals that can no longer be reduced by symmetry relations of the external and loop momenta, we employ Integration-By-Part (IBP) identities~\cite{Tkachov1981,Chetyrkin1981,Laporta:2001dd} by virtue of the Laporta algorithm~\cite{Laporta:2001dd} to reduce our integrals to a set of so-called master integrals. This step represents one of the most challenging portions of our computation due to the high degree of complexity of the integrals involved in this problem. Specifically, the integral reduction code {\tt Blade}~\cite{Guan:2024byi} developed by one of the authors incorporates the cutting edge of IBP reduction technology enabling us to achieve our results. This includes the block-triangular form~\cite{Liu:2018dmc,Guan:2019bcx}, improved seeding~\cite{Guan:2024byi,Driesse:2024xad}, the spanning-sector reduction algorithm~\cite{Guan:2024byi},  and the finite field technique~\cite{vonManteuffel:2014ixa} implemented within {\tt FiniteFlow}~\cite{Peraro:2019svx}.

To compute the residual set of master integrals we make use of the method of differential equations~\cite{Kotikov:1990kg,Kotikov:1991hm,Kotikov:1991pm,Gehrmann:1999as,Henn:2013pwa} and relate them to a basis of so-called canonical master integrals~\cite{Henn:2013pwa}. In particular, a large portion of canonical three-loop integrals was identified in refs.~\cite{Gehrmann:2024tds,DiVita:2014pza,Canko:2021xmn,Henn:2023vbd,Syrrakos:2023mor} and we relate our master integrals to the canonical results in the literature. In addition, we identified canonical one-loop and two-loop integrals as well as master integrals in two additional three-loop families that we could not identify in the literature. We checked the existing results for canonical differential equations. For the remaining master integrals we find a canonical set of master integrals using algorithmic methods~\cite{Lee:2016bib}. 
We compute boundary conditions for the differential equations using the methods of for example refs.~\cite{Henn:2020lye,Dulat:2014mda,Henn:2013woa} and confirm existing results in the decay region. We solve the canonical master integrals in terms of Chen iterated integrals~\cite{Chen:1977oja} and insert them into our newly computed scattering amplitudes in the decay region. Next, we rewrite the remaining iterated integrals in terms of multiple polylogarithms~\cite{Goncharov:2001iea} using methods outlined for example in refs.~\cite{Panzer:2014caa,Panzer:2015ida,Maitre:2005uu,Duhr:2019tlz,Duhr:2011zq,Duhr:2012fh}. The functions appearing in these scattering amplitudes were studied in some detail in ref.~\cite{Gehrmann:2024tds}. The multiple polylogarithms can readily be evaluated using public software, see for example ref.~\cite{Bauer2000,zhenjiempl}. 
To obtain results for amplitudes in the scattering region we cross the external particles accordingly and analytically continue our functions.

\section{Finite Amplitudes - UV and IR Subtraction}
\label{sec:finiteamplitudes}

The infrared and ultraviolet singularities of our scattering amplitudes are manifested in terms of poles in the dimensional regulator $\epsilon$.
The poles of an amplitude at $n^{th}$ order are described by the following universal formula in terms of its lower order counterparts such that we can define a finite remainder $\mathcal{A}_X^F$.

\bea
\label{eq:AFdef}
\mathcal{A}_X^F&&={\mathbf Z}(\alpha_S(\mu^2),\{p_i\})\mathcal{A}_X(\alpha_S,\,C^0)\nonumber\\
&&={\mathbf Z}(\alpha_S(\mu^2),\{p_i\})\mathcal{A}_X({\mathbf Z}_{\alpha_S}N_\epsilon\alpha_S^R,\,{\mathbf Z}_{C^0}C^{0,R}).
\eea

Above, $\alpha_S^R$ and $C^{0,R}$ are the renormalized strong coupling and Wilson coefficient respectively, the operator ${\mathbf Z}_{\alpha_S}$ and ${\mathbf Z}_{C^0}$ implement the usual $\overline{\text{MS}}$ ultraviolet renormalization of the strong coupling constant and the effective operator of eq.~\eqref{eq:heft} using the QCD $\beta $-function~\cite{Baikov:2016tgj,Herzog:2017ohr,Czakon:2004bu,vanRitbergen:1997va,Larin:1993tp,Tarasov:1980au} and the infrared singularities are subtracted by the universal factor ${\mathbf Z}(\alpha_S(\mu^2),\{p_i\})$~\cite{Almelid:2015jia,Aybat:2006mz,Aybat:2006wq,Catani:1998bh,Dixon:2008gr,Korchemsky:1987wg,Sterman:2002qn,Becher:2019avh}.
\beq
\label{eq:Zexp}
{\mathbf Z}(\alpha_S(\mu^2),\{p_i\},\epsilon) = \mathcal{P} e^{\-\frac{1}{4}\int_{0}^{\mu^2} \frac{d\mu^{\prime 2}}{\mu^{\prime 2}} {\mathbf \Gamma(\alpha_S(\mu^{\prime 2}),\{p_i \}, \epsilon)} }\,,
\eeq
with
\bea
\label{eq:GammaSoftDef}
 {\mathbf \Gamma(\alpha_S(\mu^2),\{p_i \}, \epsilon)}&& = \sum_{i\neq j} {\mathbf T}_i^a {\mathbf T}_j^a \Gamma_{\text{cusp}}(\alpha_S(\mu^2)) \log \frac{-s_{ij}}{\mu^2} \nonumber\\ 
 &&+ \frac{1}{2}\sum_i \mathbf{\mathds{1}} \gamma^{R_i}_c+\bold{\Delta}(\alpha_S(\mu^2),\{p_i \}).
\eea
Above, $\Gamma_\text{cusp}$ refers to the cusp anomalous dimension~\cite{Korchemsky:1987wg}, which is currently known exactly through four-loop order~\cite{Henn:2019swt,vonManteuffel:2020vjv}, and approximately at five loops \cite{Herzog:2018kwj}. 
Furthermore, $\gamma_c^R$ is the collinear anomalous dimension, obtained through four-loop order in refs.~\cite{Agarwal:2021zft,vonManteuffel:2020vjv}. The above formula was derived and calculated through three-loop order in ref.~\cite{Almelid:2015jia} and verified in $\mathcal{N}=4$ super Yang-Mills theory~\cite{Henn:2016jdu} and QCD~\cite{Caola:2022dfa,Caola:2021izf,Caola:2021rqz}. 
In ref.~\cite{Becher:2019avh}, its general structure was determined to four-loop order.
The term $\bold{\Delta}(\alpha_S(\mu^2),\{p_i \})$ is known as the correction of the dipole formula and starts at the three-loop order. 
We refer, for example, to Section 5 of ref.~\cite{Herzog:2023sgb} for further details.

We use the above definition of finite amplitudes to create a finite interference with the tree level amplitude.
\beq
\mathcal{M}_X^{F,\,(n)}=\sum\limits_{\text{spin, color}}\mathcal{A}_X^{F,\,(n)}\mathcal{A}_X^{\dagger\,F,\,(0)}.
\eeq
Finite interfered amplitudes like in the above equation are the ingredients directly feeding into the computation of scattering cross sections. 
We provide the finite interfered amplitudes for all external particle configurations in the scattering and in the decay region in ancillary files together with the arXiv submission of this Letter.

\begin{figure}[!t]
      \begin{center}
    \includegraphics[scale=.9]{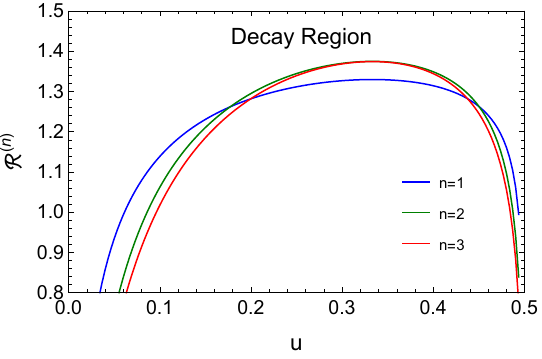}
      \end{center}
      \caption{\label{fig:Reuclideanplot}The ratio $\mathcal{R}^{(n)}$ as defined in eq.~\eqref{eq:Rdef} for $n=1$ (blue), $n=2$ (green) and $n=3$ (red) in the decay region for $a_S=0.118/\pi$.
      }
    \end{figure}

  Figure~\ref{fig:Reuclideanplot} shows the ratio $\mathcal{R}^{(n)}$,
\beq
\label{eq:Rdef}
\mathcal{R}^{(n)}=\sum_{i=0}^n a_S^i \mathbf{Re} \left[ \frac{\mathcal{M}^{F,\,(i)}_{H\to ggg}}{\mathcal{M}^{F,\,(0)}_{H\to ggg}}\right],
\eeq
 for different orders in perturbation theory in the decay region for the line $q^2=1$ and $t=u$ as a function of $u$. The central region of the plot is away from any infrared logarithmic enhancement and displays excellent perturbative stability for three-loop corrections.

\vspace{-0.5cm}
\section{Validation}
\label{sec:validation}

We performed a range of validations to ensure that our computation is correct.
\begin{enumerate}
    \item Eq.~\eqref{eq:AFdef} describes the poles in the dimensional regulator of an amplitude based on lower loop amplitudes. 
    The fact that our newly computed three-loop amplitudes satisfy this structure is a very stringent test.
    \item The two-particle collinear limit of scattering amplitudes is described by universal splitting amplitudes, which were recently computed in ref.~\cite{Guan:2024hlf} to third loop order and we validate our results in this limit.
    \item The limit of scattering amplitudes of one gluon having almost vanishing energy is described by a universal soft current, which was recently computed in ref.~\cite{Herzog:2023sgb,Chen:2023hmk} to third loop order and we validate our results in this limit.
    \item We find that the leading transcendental part of the $\mathcal{M}^{(3)}_{H\to ggg}$ amplitude corresponds~\cite{Kotikov:2001sc} to the so-called $\text{tr}(\phi^2)$ form-factor in $\mathcal{N}=4$ super Yang-Mills theory~\cite{Brandhuber:2012vm,Dixon:2020bbt,Dixon:2022rse,Lin:2021qol}. We reproduce this form factor based on an integrand from ref.~\cite{Lin:2021qol} through three-loop order and confirm this remarkable correspondence between QCD and $\mathcal{N}=4$ SYM theory for the first time at this loop order.
    \item We performed numerical evaluations of our scattering amplitudes using the auxiliary mass flow method~\cite{Liu:2017jxz,Liu:2021wks,Liu:2022mfb} implemented in 
    {\tt AMFlow}~\cite{Liu:2022chg} in the decay region and find agreement with our analytic computation in terms of multiple polylogarithms.
\end{enumerate}

\section{Conclusions}
\label{sec:conclusions}
In this Letter, we computed for the first time the scattering amplitudes for the decay of a Higgs boson to three partons as well as for the production of a Higgs boson alongside another parton in the scattering of two partons through third loop order in the generalized leading color approximation. 
These amplitudes are essential ingredients to key collider observables studying the Higgs boson at the forefront of precision phenomenology. 
Our result is the first computation of a four-particle scattering amplitude at third loop order in QCD involving a massive external particle and non-planar diagrams.
We attach our results in electronically readable files together with the arXiv submission of this Letter.

\begin{acknowledgments}
	\sect{Acknowledgments}
	We would like to thank Lance Dixon, Yan-Qing Ma and Zhen-Jie Li for useful discussions. XG, BM are supported by the United States Department of Energy, Contract DE-AC02-76SF00515. XC is supported by the Swiss National Science Foundation (SNF) under contract 200020\_219367.
    {\tt JaxoDraw}~\cite{Binosi:2003yf} was used to generate Feynman diagrams.
\end{acknowledgments}

%

\end{document}